\documentclass[pra,twocolumn,
superscriptaddress,
longbibliography,
showpacs,
aps
]{revtex4-2}
\usepackage{graphicx}
\usepackage{bm}
\usepackage{soul,xcolor}
\usepackage{physics}
\usepackage{amsmath}
\usepackage{siunitx}
\usepackage{tabularx}
\usepackage{float}
\usepackage{booktabs}
\usepackage{array,multirow}
\usepackage{hyperref}
\usepackage{orcidlink}

\hypersetup{
    colorlinks=true,
    linkcolor=blue,
    citecolor=blue,
    filecolor=blue,
    urlcolor=blue
}

\newcolumntype{Y}{>{\raggedright\arraybackslash}X}

\def\transred{{$^1$S$_0$}\,$\rightarrow$\,{$^3$P$_1$}~}

\begin{document}

\preprint{APS/123-QED}

\title{Quasi-continuous sub-$\mu$K strontium source without a high-finesse cavity stabilized laser}

\author{Sana Boughdachi\,\orcidlink{0009-0004-6305-2438}}
\thanks{These authors contributed equally to this work.}
\affiliation{Van der Waals-Zeeman Institute, Institute of Physics, University of Amsterdam, Science Park 904, 1098 XH Amsterdam, The Netherlands}
\affiliation{Toptica Photonics SE, Lochhamer Schlag 19, 82166 Gr\"afelfing, Germany}

\author{Benedikt Heizenreder\,\orcidlink{0000-0001-5611-4144}}
\thanks{These authors contributed equally to this work.}
 \affiliation{Van der Waals-Zeeman Institute, Institute of Physics, University of Amsterdam, Science Park 904, 1098 XH Amsterdam, The Netherlands}

\author{Ananya Sitaram\,\orcidlink{0000-0002-6548-4515}}
 \affiliation{Van der Waals-Zeeman Institute, Institute of Physics, University of Amsterdam, Science Park 904, 1098 XH Amsterdam, The Netherlands}

\author{Erik Dierikx\,\orcidlink{0000-0001-8061-6603}}
\affiliation{VSL National Metrology Institute, Thijssweg 11, 2629 JA Delft, the Netherlands}

\author{Yan Xie\,\orcidlink{0009-0000-8349-2888}}
\affiliation{VSL National Metrology Institute, Thijssweg 11, 2629 JA Delft, the Netherlands}

\author{Sander Klemann\,\orcidlink{0009-0005-5102-4285}}
\affiliation{SURF, Moreelsepark 48, 3511 EP Utrecht, The Netherlands}

\author{Paul Klop\,\orcidlink{0009-0008-2435-4539}}
\affiliation{SURF, Moreelsepark 48, 3511 EP Utrecht, The Netherlands}

\author{Jeroen Koelemeij\,\orcidlink{0000-0003-1005-0231}}
\affiliation{LaserLaB, Department of Physics and Astronomy, Vrije Universiteit Amsterdam, De Boelelaan 1081, 1081 HV Amsterdam, the Netherlands}

\author{Rafa\l{} Wilk\,\orcidlink{0009-0006-8393-4088}}
\affiliation{Toptica Photonics SE, Lochhamer Schlag 19, 82166 Gr\"afelfing, Germany}

\author{Florian Schreck\,\orcidlink{0000-0001-8225-8803}}
 \affiliation{Van der Waals-Zeeman Institute, Institute of Physics, University of Amsterdam, Science Park 904, 1098 XH Amsterdam, The Netherlands}
  \affiliation{QuSoft, Science Park 123, 1098 XG Amsterdam, The Netherlands}

\author{Andreas Brodschelm\,\orcidlink{0009-0000-4293-2031}}
\email{NarrowLineComb@strontiumBEC.com} 
\affiliation{Toptica Photonics SE, Lochhamer Schlag 19, 82166 Gr\"afelfing, Germany}

\date{\today}

\begin{abstract}
We demonstrate a quasi-continuous sub-\(\mu\)K strontium source achieved without the use of a high-finesse cavity-locked laser. Our frequency reference is based on a dispersion-optimized, fiber-based frequency comb that enables sub-kHz linewidths. The long-term stability of the comb is defined by an external RF reference: either a 10\,MHz RF signal from the Dutch Metrology Institute (VSL), or a tunable RF source whose long-term stability is maintained by monitoring and stabilizing the position of a narrow-line magneto-optical trap (MOT)~\cite{MOT_pos_Ytterbium_Sillus,HeizenrederMOTSpec2025}. The comb-stabilized system is benchmarked against a conventional cavity-locked laser and achieves comparable performance in broadband and single-frequency MOTs using the narrow \transred laser cooling transition. We generate high-flux, sub-\(\mu\)K samples of all three bosonic strontium isotopes and demonstrate quasi-continuous outcoupling from the MOT. These results highlight the system's suitability for compact, robust, and field-deployable continuous cold atom devices.
\end{abstract}

\maketitle           

\section{Introduction}
\label{sec:intro}
The development of ultra-cold atomic sources has revolutionized precision measurement, driving progress in optical lattice clocks~\cite{bothwell_resolving_2022,katori_ultrastable_2003,mcgrew_atomic_2018} and other quantum sensors~\cite{Atom_Interferometry_Holger, panda_coherence_2024,stray_quantum_2022}. In recent years, there has been growing interest in continuous high-density ultra-cold atomic sources~\cite{chen_continuous_2022,Modeling_swadheen,schafer_continuous_2025,Steady_state_magneto_optical_Escudero,Singh2022det,crossed_moving_optical_lattices_Okaba,Atom_computing,Li2025_fast_continuous_coherent,Chiu2025_continuous_3000_qubit}. Strontium, as an alkaline-earth element, is particularly well-suited for continuous cold atom sources due to its narrow optical transitions, which enable narrow-linewidth laser cooling schemes~\cite{Katori_Photon_Recoil_Temperature,Chen2025_narrowline_sisyphus,Chen2019_sisyphus_lattice_decelerator,Hobson2020_midIR_MOT_Sr,Cooper2018_sideband_cooling_Sr_tweezer}. Traditionally, such narrow-line cooling schemes rely on laser systems stabilized to high-finesse optical cavities~\cite{StellmerPhD2013,qiao_ultrastable_2019}, which can provide laser linewidths well below that of the narrow cooling transition. While these cavities offer excellent short-term frequency stability, they require meticulous alignment and are sensitive to environmental perturbations that limit operational robustness and long-term reliability. Thus, a key challenge is to achieve narrow-linewidth laser stabilization without reliance on environmentally sensitive cavities.
One alternative stabilization method is via a frequency comb, though typical fiber-based combs suffer from linewidth broadening due to oscillator pump-induced noise, environmental fluctuations, and quantum noise, compromising their use in narrow-linewidth laser cooling or other precision applications~\cite{newbury2007low}. 

To overcome these limitations, we introduce a dispersion-engineered frequency comb and demonstrate its use for a quasi-continuous ultracold strontium source. 
The frequency comb is generated using mode-locked Er:fiber oscillators, which are intrinsically insensitive to pump-induced fluctuations by dispersion engineering. 
These measures preserve the coherence and spectral purity of individual comb teeth, enabling direct laser stabilization with sub-kHz precision by locking its repetition-rate to an RF reference. 
The reduced susceptibility to thermal and vibrational noise also enables easier integration into compact, transportable quantum platforms. 
Through a non-linear conversion processes, a broad spectral range extending from 420\,nm to beyond 2000\,nm can be obtained, creating an architecture adaptable for narrow-linewidth laser cooling of strontium at 435\,THz but also for a wide range of other narrow-linewidth alkaline-earth(-like)~\cite{adamczyk_two-photon_2025,Clock-Line-Mediated_Chun-Chia} and lanthanide atoms~\cite{erbium_Frisch,dysprosium_Mingwu}. 
We use our fully fiber-based frequency comb to generate the ultracold strontium source by directly locking the narrow-line cooling laser to it without any need of a high-finesse free-space reference cavity. 
Thanks to its broad coverage and flexible offset-locking capabilities, our system supports laser cooling of all three stable bosonic strontium isotopes using a single unified laser architecture. Combined with a robust laser cooling setup, this enables the generation of a continuous source of ultra-cold strontium atoms at \(\mu\)K temperatures. 
To ensure long-term frequency stability of the comb, its repetition rate must be stabilized to an external reference. 
We employ two approaches, which we compare against a cavity-based laser system: a highly stable external RF reference, specifically a 10\,MHz signal provided by the Dutch Metrology Institute VSL \cite{dierikx_white_2016}, and a self-contained stabilization scheme locking the comb to a tunable oven-controlled crystal oscillator, which is referenced to the MOT position \cite{HeizenrederMOTSpec2025}. The scheme enables robust operation even in the absence of external frequency standards.
Finally, we introduce a technique for quasi-continuous out-coupling of sub-\(\mu\)K atoms from the MOT, highlighting the potential of this platform for zero-dead-time optical lattice clocks~\cite{katori_longitudinal_2021,Zero-Dead-Time_Operation_Biedermann}, portable clocks~\cite{portable_optical_clock_Grotti,takamoto_test_2020}, and quantum devices that require quasi-continuous atomic sources~\cite{Atom_computing,Continuous_Gyger,chen_continuous_2022,Li2025_fast_continuous_coherent,Chiu2025_continuous_3000_qubit}.

The remainder of this article is organized as follows: in Section~\ref{sec:exp_setup:noise}, we give some background on the fundamental noise sources of frequency combs and the theoretical framework to model them.
We then characterize the influence of pump-induced noise on the repetition rate of our frequency comb, validate the measurements against theory predictions~(\ref{sec:exp_setup:short}), and discuss long-term stability of the repetition rate and carrier-envelope offset frequency~(\ref{sec:exp_setup:long}).
We next show how the frequency comb can be used as laser frequency reference for our quasi-continuous ultracold strontium source (\ref{sec:exp_setup:source}), characterize the long-term stability of the system when using either a stable 10~MHz reference or the MOT position to stabilize the frequency comb~(\ref{sec:exp_setup:RF}), and benchmark the performance against that of a conventional cavity-locked laser system.
Finally, we demonstrate how we can use this architecture to achieve quasi-continuous outcoupling of atoms from a MOT~(\ref{sec:exp_setup:quasi-continuous}).

\section{Cavity-Free Frequency Reference}
\label{sec:exp_setup:comb}
\raggedbottom

Optical frequency combs (OFCs) produce a stable and evenly spaced set of optical frequencies, effectively bridging the optical and radio frequency domains. Each comb line is described by the relation:

\begin{align}
f_n = f_{\text{CEO}} + n f_{\text{r}},
\end{align}

where \( f_{\text{CEO}} \) is the carrier-envelope offset frequency, \( f_{\text{r}} \) is the repetition rate, and \( n \) is an integer mode index ~\cite{kliese2016difference}.

One key advantage of OFCs, particularly those generated by Er-doped fiber-based oscillators, is their stability and ease of use. 
In such OFCs, the Er-fiber oscillator locked to the RF reference typically operates at 1550~nm. 
The 80~MHz output of the Er-doped fiber oscillator is amplified in an Er-doped fiber and then spectrally broadened through self-phase modulation (SPM) in a highly nonlinear fiber (HNLF). The broadened spectrum is subsequently frequency-doubled using second-harmonic generation (SHG) to reach the necessary spectral range. 
For our application, a continuous-wave (CW) diode laser operating at 689~nm is then phase-locked to the corresponding comb tooth to provide sufficient optical power.
This provides an intense single-line source at the spectral position needed for cooling of strontium on the 7.5~kHz-wide \transred intercombination transition. 
Standard fiber-based RF-locked frequency combs typically exhibit intrinsic linewidths greater than 20\,kHz \cite{newbury2006reducing}.
However, to reach recoil-limited performance on the intercombination line~\cite{Katori_Photon_Recoil_Temperature}, we strive to have a laser linewidth less than that of the natural linewidth of the transition. We use a concept developed in our previous research collaboration~\cite{hutter2023femtosecond} and optimize its design to achieve exceptionally narrow linewidth operation at wavelengths required for optical cooling applications.

\subsection{FUNDAMENTAL NOISE SOURCES}
\label{sec:exp_setup:noise}
There are several noise sources that fundamentally limit the linewidth of a frequency comb.  
Environmental disturbances such as thermal fluctuations, acoustic noise, and mechanical vibrations contribute to slow drifts and frequency jitter in the acoustic frequency range. Quantum noise arising from spontaneous emission and shot noise impose fundamental limits on coherence. Furthermore, fluctuations in the pump laser power introduce pump-induced noise, which, along with other environmental and quantum noise sources, can couple into both the repetition rate and the carrier-envelope offset frequency of the oscillator \cite{newbury2007low,hutter2023femtosecond, paschotta2006optical}.

To understand how different noise sources contribute to the linewidth of individual comb lines, the elastic tape model provides a valuable conceptual and mathematical framework ~{\cite{washburn2005response}}. In this model, the optical frequency comb spectrum is represented as an elastic tape spanning the frequency domain, where fluctuations in system parameters lead to changes in its overall position and spacing. Specifically, shifts in the carrier-envelope offset frequency ($f_{\mathrm{CEO}}$) correspond to parallel translations of the tape, while changes in the repetition rate ($f_{\mathrm{r}}$) correspond to stretching or compression, modifying the mode spacing ~\cite{telle2002kerr}.
When a system parameter $X$ (such as pump power, cavity length, or environmental conditions) fluctuates, it perturbs both the repetition rate $f_{\mathrm{r}}$ and the carrier-envelope offset frequency $f_{\mathrm{CEO}}$. The resulting frequency shift depends on the location of the tooth relative to a specific point in the spectrum, known as the \textit{fixed point}.
For a given perturbation \( X \), the mode index of the associated fixed point is given by ~\cite{washburn2005response}:

\begin{align}
n^{X}_{\text{fix}} = -\frac{\partial f_{\text{CEO}} / \partial X}{\partial f_{\text{r}} / \partial X},
\end{align}

and the corresponding fixed point frequency is:

\begin{align}
f^{X}_{\text{fix}} = n^{X}_{\text{fix}} \cdot f_{\text{r}} + f_{\text{CEO}}.
\end{align}

At the fixed point, the frequency shift caused by fluctuations in \( X \) is minimized due to the cancellation of contributions from \( \Delta f_{\text{r}} \) and \( \Delta f_{\text{CEO}} \). Comb lines located farther from this point exhibit increased sensitivity and larger fluctuations. 

It is also important to consider the physical nature of the perturbation. If the disturbance primarily alters the optical path length, and therefore changes the pulse round-trip time without significantly affecting the carrier phase, the fixed point appears near the carrier frequency. In this case, comb modes around the carrier frequency remain relatively stable, while low-frequency modes (such as those near zero frequency) show larger shifts. Conversely, if the perturbation equally affects the carrier phase and pulse timing, the fixed point moves toward zero frequency. Here, the low-frequency comb modes are more stable, and modes around the carrier frequency exhibit larger fluctuations. This further emphasizes that each noise source contributes its own characteristic fluctuations and is associated with a distinct fixed point, determined by the nature of its coupling to the comb parameters \cite{newbury2007low}. The total frequency noise power spectral density (PSD), \( S_{\Delta\nu,\nu}(f) \), is given by~\cite{newbury2007low}:

\begin{align} \label{Equation 1}
S_{\Delta\nu,\nu}(f) = \notag \\
f_{\text{r}}^2 \Big[ &
S_{\text{rep}}^{\text{quant}}(f)(\nu - \nu_c)^2 \notag \\
& + S_{\text{rep}}^{\text{pump}}(f)(\nu - \nu_{\text{fix,pump}})^2 \notag \\
& + S_{\text{rep}}^{\text{env}}(f)\nu^2 \Big],
\end{align}

\noindent where each term represents the noise contribution from a particular source: quantum fluctuations, pump power noise, and environmental disturbances. The quadratic dependence on $\nu-\nu_{\text{fix}}$ highlights that the magnitude of noise increases with distance from the respective reference point. In this model, \( \nu_c \) represents the optical carrier frequency of the mode-locked oscillator (193~THz for a 1550~nm system). Quantum-limited noise has minimal impact on the repetition rate at this spectral position. 
The spectral location of the pump-induced fixed point \( \nu_{\text{fix, pump}} \) is not constant; rather, it is dictated by the oscillator design and can be tuned by modifying intracavity parameters.
Environmental noise affects both the phase and group velocity equally, causing its fixed point to be at the origin of the frequency scale, while often dominant at low frequencies, can be effectively mitigated through mechanical and thermal isolation strategies.

Quantum noise is a broadband fluctuation that sets a lower bound on comb stability and predominantly perturbs the pulse timing, and thus the repetition rate, rather than the carrier phase of the pulse \cite{liao2020dual}.
Pump-induced noise, however, plays a dominant role in fiber-based oscillators. Due to its strong coupling with gain dynamics, it introduces substantial fluctuations in both the repetition rate \( f_{\text{r}} \) and carrier envelope offset frequency \( f_{\text{CEO}} \), significantly broadening the comb linewidth, especially at spectral positions far from the pump noise fixed point.

The sensitivity of the repetition rate to pump power, \( \frac{df_{\text{r}}}{dP} \), originates from the combined effects of intracavity dispersion, nonlinear dynamics, and gain contribution within the laser cavity. These effects include spectral shifts due to changes in the pulse center frequency, third-order dispersion (TOD), resonant gain contribution in the erbium-doped fiber, and self-steepening effect (SSD). Together, they define the following expression for the pump-induced repetition rate sensitivity~\cite{newbury2007low}:

\begin{align} \label{Equation 2}
\frac{df_{\text{r}}}{dP} = -f_{\text{r}}^2 \bigg(
\underbrace{\beta_{2,\text{cav}} \frac{d\omega_\text{c}}{dP}}_{\textcolor{black}{\text{Spectral shift}}}
+ \underbrace{\frac{\beta_3 \omega_{\text{rms}}^2}{2P}}_{\textcolor{black}{\text{TOD}}} \notag \\
+ \underbrace{\frac{\nu_{3\text{dB}}^{\text{Er}}}{3P \nu_{3\text{dB}} \Omega_\text{g}}}_{\textcolor{black}{\text{Resonant gain}}}
+ \underbrace{\frac{3 \mu \gamma A^2}{2P \omega_0}}_{\textcolor{black}{\text{SSD}}} \bigg)
\end{align}

The second-order dispersion \( \beta_2 \) and third-order dispersion \( \beta_3 \) represent the lumped dispersion accumulated over all fiber segments in the cavity~\cite{newbury2005theory}. Notably, the first term, involving \( \beta_2 \), captures the sensitivity of the repetition rate to shifts in the pulse’s central frequency \( \omega_c \), which themselves arise from pump-induced changes in the intracavity gain contribution. The second term accounts for the contribution from third-order dispersion where \( \omega_\text{rms} \) is the spectral width of the oscillator's optical output spectrum. The third arises from resonant gain contribution due to the finite gain bandwidth of the erbium-doped fiber, and the fourth term originates from self-steepening caused by nonlinear pulse propagation with peak intensity \(A^2\).

The location of the fixed point depends on both the sensitivity of the repetition rate and the response of the intracavity phase to the same perturbation. 
The spectral position of the pump-induced fixed point can be derived from the relative sensitivities of the carrier-envelope phase \( \phi \) and the repetition rate \( f_{\text{r}} \) to changes in pump power. This frequency, denoted \( \nu_{\text{fix}}^{\text{pump}} \), is given by

\begin{align} \label{Equation 3}
\nu_{\text{fix}}^{\text{pump}} = \nu_c + f_{\text{r}}^2 \left( \frac{d\phi / dP}{df_{\text{r}} / dP} \right),
\end{align}

\noindent where \( \nu_c \) is the optical carrier frequency, \( d\phi/dP \) is the sensitivity of the carrier phase to pump power \cite{newbury2007low}. Once the spectral position of the fixed point is determined, it becomes possible to evaluate how strongly a given comb line is affected by pump-induced noise. A practical measure of this influence is the optical linewidth, which can be directly linked to the frequency noise power spectral density. At a given frequency \( \nu \), the linewidth \( \delta_\nu \) can be estimated using the relation \cite{newbury2007low}

\begin{align} \label{Equation 4}
\delta_\nu = \pi \sqrt{S_{\nu}(0) f_{3\text{dB}}},
\end{align}
where \( S_{\nu}(0) \) is the low-frequency (near-zero offset) value of the frequency noise PSD at the target optical frequency \( \nu \), and \( f_{3\text{dB}} \) is the 3-dB bandwidth of the frequency noise spectrum.

\subsection{SHORT-TERM STABILITY}
\label{sec:exp_setup:short}
To address these noise sensitivities and achieve stable frequency operation, we implement a frequency comb based on a mode-locked Er-doped fiber oscillator, which offers passive, robust, and environmentally resilient pulse formation~\cite{Fermann90}. The oscillator cavity incorporates two types of fibers to manage dispersion: erbium-doped gain fiber (EDF), which exhibits normal second-order dispersion, and passive polarization-maintaining (PM) fiber, which exhibits anomalous dispersion. By adjusting the length ratio between these two fiber segments, the net intracavity dispersion \( \beta_{2,\text{cav}} \) can be precisely controlled. This architecture enables sub-kHz passive frequency stability over a broad spectral range, as demonstrated in similar designs~\cite{hutter2023femtosecond}, allowing for optimization of frequency stability while keeping other laser parameters fixed.

To validate the theoretical predictions and quantify the influence of pump-induced noise on the repetition rate, we experimentally measured the sensitivity of the repetition rate on pump power fluctuations (\( \frac{df_{\text{r}}}{dP} \)). A modulation was applied to the continuous-wave (CW) pump power. The resulting modulation in the oscillator's repetition frequency \( f_{\text{r}} \) was monitored using a frequency counter.

For \( f_{\mathrm{CEO}} \) detection, we employ a fiber-coupled \( f\)-to-\(2f \) interferometer similar to the approach demonstrated in~\cite{newbury2007operation}, which offers robust alignment and full fiber integration. Although other implementations (e.g. free-space setups) are possible, the fiber-based configuration supports long-term stability and integration in compact platforms.

\begin{figure}[]
    \centering
    \includegraphics[width=8.5cm]{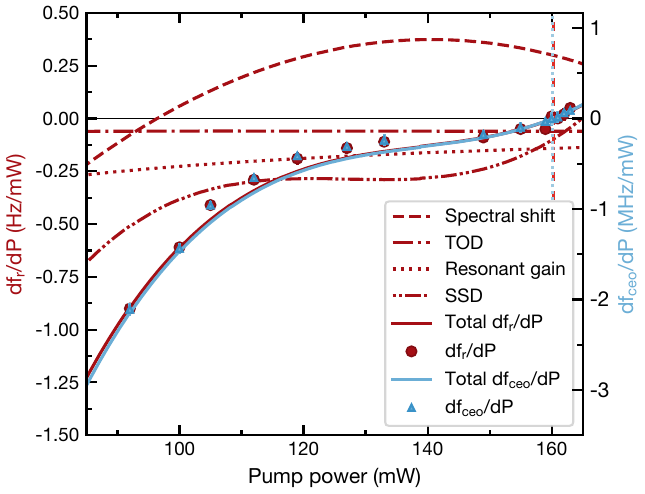}
    \caption[Sensitivity of Repetition Frequency to Pump Power]
    {\label{fig:df_dP_sensitivity} Measured and theoretical sensitivity of the repetition rate \( \frac{df_{\mathrm{r}}}{dP} \) (left axis) and carrier-envelope offset frequency \( \frac{df_{\mathrm{CEO}}}{dP} \) (right axis) as a function of pump power. Experimental data for \( \frac{df_{\mathrm{r}}}{dP} \) (red points) is compared with theoretical contributions from spectral shift (dashed), third-order dispersion (TOD, dot-dashed), resonant gain (doted), and self-steepening (SSD, dot-dot-dashed), with their total shown as a solid red line. Measured \( \frac{df_{\mathrm{CEO}}}{dP} \) values are shown as blue triangle. The theoretical repetition rate sensitivity crosses zero at a pump power of \text{160.27\,mW} (vertical red dashed line), while the theoretical carrier-envelope offset sensitivity crosses zero at \text{160.21\,mW} (vertical blue dashed line).
    }
\end{figure}

\begin{table}[b] 
\centering
\caption{Values of laser parameters used in Eq. \ref{Equation 2} for the modeled repetition rate sensitivity to pump power.}
\begin{tabular}{|c|l|c|l|}
\hline
\textbf{Symbol} & \textbf{Quantity} & \textbf{Value} & \textbf{Ref.} \\ \hline
$f_\text{r}$ & Repetition rate frequency & 80 MHz & \\ \hline
$\nu_{3\text{dB}}^{\text{Er}}$ & Erbium response bandwidth & 1.6 kHz & \cite{washburn2005response} \\ \hline
$\nu_{3\text{dB}}$ & Laser response bandwidth & 12 kHz &\\ \hline
$\Omega_\text{g}$ & Gain spectrum bandwidth & $0.3 \, \text{THz}$ & \cite{washburn2005response} \\ \hline
$\mu$ & Self-steepening factor & 1.3 & \cite{washburn2005response} \\ \hline
$\gamma$ & Lumped nonlinearity & $4 \, \text{kW}^{-1}$ &  \\ \hline
$\omega_\text{rms}$ & Root Mean Square frequency & ${2\pi c \Delta\lambda_{\text{rms}}}/({\lambda_c^2})$ & \cite{newbury2005theory}\\ \hline
$\omega_0$ & Gain peak frequency & $2\pi c / 1560 \, \text{nm}$ & \\ \hline
\end{tabular}
\label{tab:laser_parameters}
\end{table}

Figure~\ref{fig:df_dP_sensitivity} presents both the measured and modeled sensitivity of the repetition rate \( df_{\text{r}}/dP \) and the carrier-envelope offset frequency \( df_{\text{CEO}}/dP \) to pump power variations. 
To model the sensitivity, we use Eq.~\ref{Equation 2}, with the values for the parameters as listed in Table ~\ref{tab:laser_parameters}. 
The experimental data for \( df_{\text{r}}/dP \) (red points) shows a nonlinear dependence on $P$, starting from approximately \(-1.3~\text{Hz/mW}\) at lower pump powers and approaching zero near \(160~\text{mW}\). This zero crossing in \( df_{\text{r}}/dP \) corresponds to a singularity in the fixed point frequency, making the latter highly tunable through small changes in $P$ around this operating point.
The oscillator mode-locking window sets practical boundaries on the pump power tuning range without losing comb operation. This trend is well-captured by the theoretical model as shown in Eqn.\,\ref{Equation 2} (solid red line), which accounts for contributions from spectral shift, third-order dispersion, resonant gain, and self-steepening effects. 

Simultaneously, the measured \( df_{\text{CEO}}/dP \) values (blue triangles) reveal a similarly strong, nonlinear behavior, ranging from \(-3.5~\text{MHz/mW}\) to nearly zero. The similar behavior and different magnitudes of \( df_{\text{r}}/dP \) and \( df_{\text{CEO}}/dP \) are critical for determining the position of the pump-induced fixed point. Notably, both derivatives tend toward zero at nearly the same pump power, suggesting an optimal operating regime where pump-induced noise can be significantly minimized across the entire comb spectrum. This analysis provides a foundation for tailoring the comb's operating point to suppress linewidth broadening in a broad range of optical frequencies.
The observed correlation between \( df_{\text{r}}/dP \) and \( df_{\text{CEO}}/dP \) arises because both are influenced by the same cavity dynamics, such as gain shaping and nonlinear phase shifts. In mode-locked oscillators, a pump-induced change in the pulse envelope timing directly affects \( f_{\text{CEO}} \), and often also alters \( f_{\text{r}} \). Therefore, their sensitivities to pump power tend to follow similar trends, reflecting the underlying physical processes.

\begin{figure}[]
    \centering
    \includegraphics[width=8.5cm]{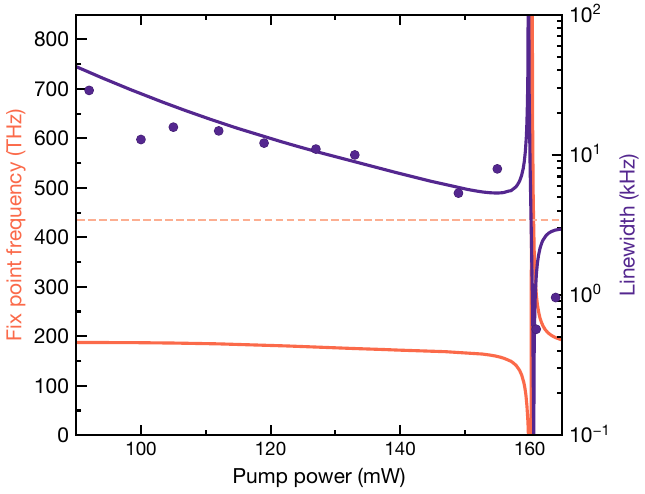}
    \caption[Pump-Induced Fixed Point Frequency Shift]
    {\label{fig:fixpoint_linewidth} Calculated and experimental results for the pump-induced fixed point frequency (left axis, orange curve) and optical linewidth at 689~nm (right axis, purple points) as a function of pump power. The orange curve shows the calculated fixed point frequency, while the purple points represent the experimentally measured linewidth values, averaged over 10 independent measurements per point. The purple line corresponds to the theoretical linewidth prediction. An optimal operating point at 161~mW can be found where the fixed point (horizontal dashed line at 435~THz) aligns with the atomic transition frequency, and the linewidth reaches a minimum below 1~kHz.
}
\end{figure}

As shown in Figure~\ref{fig:fixpoint_linewidth}, with increasing pump power, the fixed point frequency shifts considerably, reaching a value close to the target atomic transition frequency of 435~THz at approximately 161~mW.
The measured linewidth data shows a clear correlation with the fixed point behavior. At pump powers far from 161~mW, the linewidth is substantially broadened, indicating that noise contributions from the pump strongly affect the spectral purity. However, near 161~mW, the linewidth reaches a minimum below 1~kHz, suggesting that the effective noise at the transition frequency is minimized due to the proximity of the fixed point.
By selecting a pump power where the fixed point aligns with the atomic transition frequency, high spectral purity and long-term stability can be obtained without the complexity of optical cavity locking. 

Beyond identifying the fixed point, it is important to understand how frequency noise behaves as one moves away from it. As expressed in Eq.~\ref{Equation 1}, the frequency noise power spectral density \( S_{\Delta\nu,\nu}(f) \) increases quadratically with the distance from the fixed point. The pump pump-induced noise term \( S^{\text{pump}}_{\text{rep}}(f) \) reaches a minimum in the pump power region where the fixed point can be tailored. Therefore, the linewidths of the comb teeth will be drastically reduced across the comb spectrum in this operation mode. However, the linewidth of a selected comb tooth is minimized if the fixed point frequency \( \nu_{\text{fix,pump}} \) coincides with the target frequency \( \nu \).

\subsection{LONG-TERM STABILITY}
\label{sec:exp_setup:long}

Our goal is to define long-term stability by the stability of the external RF reference, which requires a robust frequency locking mechanism and the ability to maintain the pump-induced fixed point stable over time.
In our setup, the repetition rate $f_{\mathrm{r}}$ is locked by means of a piezo actuator controlling the optical path length of the cavity, while the carrier-envelope offset frequency $f_{\mathrm{CEO}}$ is stabilized by modulating the pump power.

However, locking \( f_{\mathrm{r}} \) and \( f_{\mathrm{CEO}} \) is essential to stabilize the frequency comb.
Any significant drift away from the optimal pump power alters the intra-cavity conditions, shifts the pump-induced fixed point, and increases noise at the target atomic transition. 
To address this, we ensure that the stabilization scheme maintains the pump power within a narrow window around the optimal point (e.g., 161~mW), where the fixed point aligns with 
435~THz and sub-kHz linewidths are achieved. 


\section{Continuous ultracold strontium source without a free-space cavity stabilized laser}
\label{sec:exp_setup}

In this section, we combine our dispersion-engineered frequency comb system with a continuous ultracold atom source apparatus. We begin by introducing the design and operation principles of this architecture. Next, we compare the performance of the comb-locked laser system, using different long-term frequency references, against a conventional cavity-stabilized setup. In our approach, the frequency comb provides the short-term stability typically gained by a free-space optical cavity, while long-term stability, usually derived from hot-vapor atomic spectroscopy, is achieved via an external RF reference or through a self-contained stabilization scheme based on MOT position feedback. Finally, we present a simple method for quasi-continuous outcoupling of atoms from a continuously replenished five-beam MOT.

\subsection{CONTINUOUS ULTRACOLD STRONTIUM SOURCE}
\label{sec:exp_setup:source}

\begin{figure*}[]
    \centering
    \includegraphics[scale=0.9]{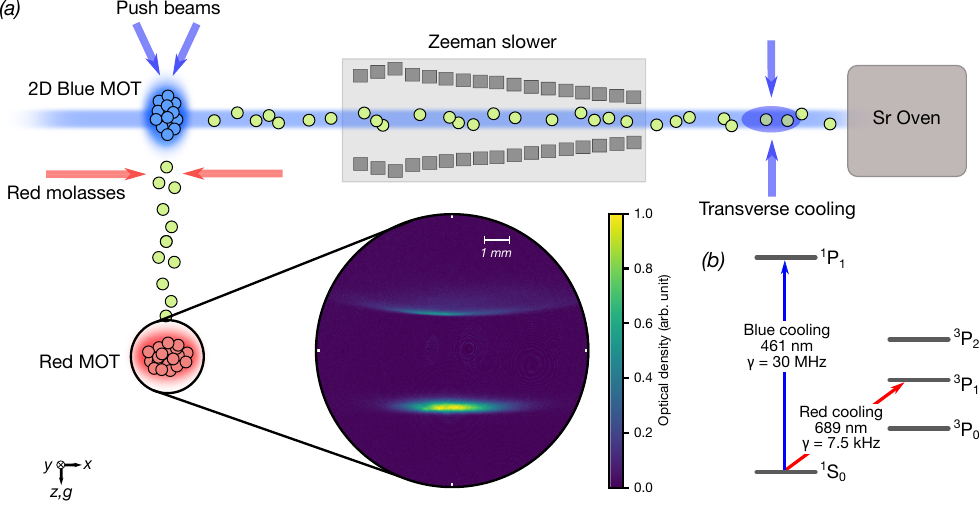}
    \caption[]
    {\label{fig: ExpSchematic} (a) Experimental schematic. Strontium atoms exit the oven and propagate in the -$\hat{x}$ direction. Transverse cooling beams operating on the $^1$S$_0\,\rightarrow\,^1$P$_1$ transition collimate the atomic beam along the $\hat{y}$ and $\hat{z}$ axes. The atoms are slowed by a Zeeman slower before arriving in a 2D blue MOT, both operating on the $^1$S$_0\,\rightarrow\,^1$P$_1$ transition. Atoms then fall from the blue MOT down into a lower chamber due to gravity, where they are captured in a 3D MOT operating on the $^1$S$_0\,\rightarrow\,^3$P$_1$ transition. Push beams on the $^1$S$_0\,\rightarrow\,^1$P$_1$ transition and red molasses beams addressing the $^1$S$_0\,\rightarrow\,^3$P$_1$ transition help to minimize loss of atoms between the blue and red MOTs. Atoms are captured in the BB (zoomed image, bottom) and SF (zoomed image, top) MOT simultaneously in a ``dual-position MOT" (see Section~\ref{sec:exp_setup:quasi-continuous}). (b) Energy level diagram for strontium showing the blue and red cooling transitions.  }
\end{figure*}

The design of our continuous cold atom apparatus is derived from previous work in our group \cite{Steady_state_magneto_optical_Escudero,chen_continuous_2022}. 
We have simplified the previous design, reducing the total volume of the vacuum system to just $0.5~$m$^3$. 
The continuous ultracold atom source is created through the simultaneous execution of a sequence of cooling stages that are separated in space, as shown in Fig.~\ref{fig: ExpSchematic}(a). The sequence starts with a hot atomic beam that effuses out of an oven heated to approximately $460\,^{\circ}\mathrm{C}$. The atomic beam is cooled along the two transverse axes using two retroreflected beams addressing the blue cooling transition ($^1$S$_0\,\rightarrow\,^1$P$_1$, see Fig.~\ref{fig: ExpSchematic}(b)). We slow the atomic beam with a permanent-magnet Zeeman slower using $\sim 200$\,mW of light 242\,MHz red-detuned from the blue transition.
The slowed atoms are captured by a 2D magneto-optical trap (MOT) operating on the broad blue cooling transition and cooled to temperatures on the order of a few mK.
The quadrupole magnetic field for the 2D MOT is produced by stacks of permanent magnets and has a gradient of $\sim$20\,G/cm. The 2D MOT laser beams are 32~MHz red-detuned from the blue cooling transition and have a beam waist of 1.5\,cm and a peak intensity $I/I_\mathrm{sat} = 0.4$, where $I_{\mathrm{sat}} = 40.3~\mathrm{mW/cm^2}$ is the saturation intensity of the blue cooling transition.
Since there is no confinement in the direction of gravity in the 2D MOT, atoms can fall freely out of the MOT and into a lower chamber. The atoms' upward movement is limited by two 461-nm ``push" beams angled symmetrically about the $-\hat{z}$-axis. 
All laser beams addressing the blue cooling transition are produced by injection-locked lasers, for which the parent laser is stabilized using modulation transfer spectroscopy from a hot Sr vapor.
As the atoms drop to the bottom chamber, the atomic beam is collimated directly below the blue 2D MOT beams, using two red molasses beams addressing the narrow red cooling transition ($^1$S$_0\,\rightarrow\,^3$P$_1$, 7.5\,kHz wide).
The average time that atoms spend in the blue MOT is less than 5\,ms, enabling high atomic flux into the lower chamber without the need for repump lasers. This is possible despite the partially open nature of the $^1$S$_0\,\rightarrow\,^1$P$_1$ transition, which allows decay into metastable states~\cite{Cooper2018_sideband_cooling_Sr_tweezer,Xinye_3D3_loss_channel}.

As atoms arrive in the bottom chamber containing a broadband five-beam red MOT (BB MOT), comprised of two counterpropagating beam pairs along the horizontal axes and one upwards propagating MOT beam.
The latter beam is modulated at a frequency of 70~kHz over a range of $1.05 - 2.55$~MHz red-detuned from the MOT transition, allowing it to also act as a white light slower~\cite{Zhu1991}.
The horizontal MOT beams are retro-reflected, with a waist of 15~mm and peak intensity of 1.1~mW/cm$^2$, or $\sim 366~I_{\mathrm{sat}}$.
They are modulated at a frequency of 30~kHz to cover a range of $1.05 - 2.4$~MHz red-detuned from the MOT transition.
The quadrupole magnetic field is produced by a quadrupole coil pair, with a gradient of approximately $0.39$\,G/cm.
Notably, only five laser beams are used to produce the red MOT. 
There is no need for a vertically downwards propagating MOT beam, as gravity provides a sufficient force to balance out the force of the upwards propagating MOT beam.
With these parameters, we achieve a steady state BB red MOT of $^{88}$Sr with $4.25\times10^8$ atoms. For further information on the apparatus design, see Ref.~\cite{ZhouPhD2024}. 

\subsection{DIFFERENT REFERENCES FOR LONG-TERM STABILITY}
\label{sec:exp_setup:RF}
In this section, we evaluate two methods for synchronizing the frequency comb and thereby providing an absolute and long-term stable optical frequency reference for the red MOT. We benchmark the two lock methods for our application by characterizing the red MOT produced with each and comparing them with a red MOT created using a more conventional laser system, stabilized on short ($<1$\,s) timescales to a cavity, and on longer timescales through hot vapor cell spectroscopy \cite{StellmerPhD2013,qiao_ultrastable_2019}.

For the first method, we use a 10~MHz RF source from the Dutch Metrology Institute (VSL) as a reference to the frequency comb. 
This RF signal is produced by a local White Rabbit switch locked to a White Rabbit grandmaster (GM) switch located at VSL Delft, which itself is referenced to UTC(VSL). The White Rabbit switches are connected through about 100~km of optical fiber of the SURF research and education network. The White Rabbit link forms part of a larger network, the SURF Time\&Frequency network, which is used to disseminate UTC(VSL) to multiple locations in the Netherlands via the White Rabbit protocol. Typical frequency stability of the provided 10~MHz is between $10^{-12}$ and $10^{-11}$ fractionally after 1~s of averaging, dropping down to $\sim 10^{-14}$ after 100~s of averaging, and dominated by white phase noise between 1 and 100~s of averaging~\cite{vsl2025wr,White_Rabbit,surf2024timefrequency, dierikx_white_2016, VanTour2017WhiteRabbit}.

In the second method, we reference the frequency comb to a quartz oscillator and actively stabilize the MOT position by adjusting the frequency of the quartz oscillator.
In this case, the comb is not referenced to any external RF signal. The quartz oscillator combined with the slow feedback loop from the MOT position determines the stability of the system~\cite{MOT_pos_Ytterbium_Sillus,HeizenrederMOTSpec2025}.
In a five-beam MOT, the vertical MOT position is directly related to the frequency of the laser light, so we measure the frequency stability of our system by examining the MOT position over time. We continuously determine the MOT position by fitting fluorescence images of the MOT. This technique can be easily adapted to enhance the long-term stability of conventional free-space cavity laser systems.

\begin{figure}[]
    \centering
    \includegraphics[width=8.5cm]{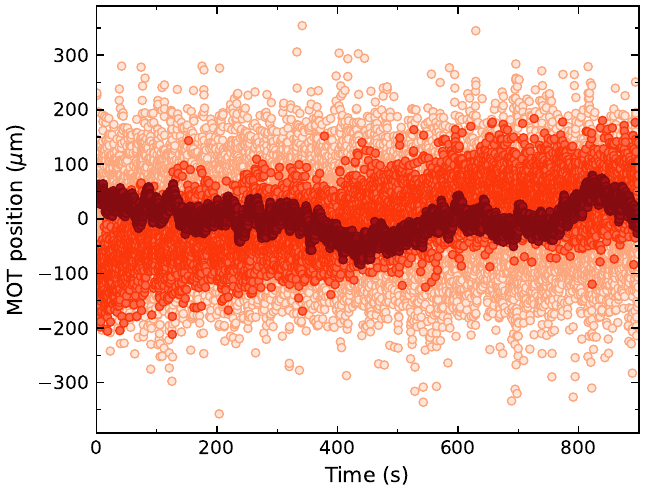}
    \caption[]
    {\label{fig: MOT stability} Stability of the position of the BB red MOT over 900\,s for the cavity-stabilized system long-term referenced to hot vapor spectroscopy (dark red dots) vs. the comb-stabilized system referenced to a 10\,MHz signal from VSL (orange dots) and to the MOT position (light orange dots). The standard position deviation for the three stabilization methods over 900\,s are $\sigma _{\mathrm{cavity}}=27.1~\mu$m, $\sigma _{\mathrm{VSL}}=60.73~\mu$m, and $\sigma _{\mathrm{active}}=83.6~\mu$m, respectively. All measurements are taken at a small magnetic field gradient ($0.39$~G/cm) to magnify the differences.}
\end{figure}

We examine the stability of the BB MOT position using the conventional laser system and using the comb laser system with each of the two comb locking methods over a period of 900~s each, as shown in Fig.~\ref{fig: MOT stability}.
While the conventional laser system provides the best short-term stability, with a standard deviation of the position of 27.1~$\mu$m over 900\,s, there are larger variations in the mean MOT position over longer time scales, but no clear drift.
With the comb locked to the VSL RF source, the standard deviation of the MOT position over 900~s is increased to 60.73~$\mu$m. A steady drift is observed over the whole measurement period, which could come from a drift of the VSL RF frequency or of the comb repetition rate locking electronics. Other reasons could be magnetic field or MOT laser intensity drift. However, those drifts would also influence the MOT position when using the cavity-locked laser system, which we don't observe.
Stabilizing the comb on the red MOT position produces a standard MOT position deviation of 83.6~$\mu$m over 900\,s. 
Long-term drift is absent as it is removed by the MOT-position lock. If the MOT fluorescence imaging system is stable, this can contribute to more reliable loading of MOT atoms into dipole traps compared to the previous two methods, rendering subsequent experimental stages more reliable. We further analyze the short-term stability by dividing the data into 400 non-overlapping intervals of 2\,s duration and computing the standard deviation of the MOT position within each interval. The reported values correspond to the mean of these interval-wise standard deviations, with the uncertainty given by their standard deviation. For the conventional laser system, we obtain $8.2 \pm 1.2\,\mu\mathrm{m}$, for the VSL $41.2 \pm 7.7\,\mu\mathrm{m}$, and for the MOT position stabilization scheme $79.2 \pm 16.5\,\mu\mathrm{m}$. For the comb-locked methods, the stability over 2\,s is primarily limited by the stability of the RF source used to synchronize the frequency comb within the same time frame. In particular, the performance of the MOT position stabilization scheme could be significantly improved by employing a quartz oscillator with enhanced short-term stability.

\begin{table}[b]
\centering
\begin{tabular}{|>{\centering\arraybackslash}p{1.75cm}|>{\centering\arraybackslash}p{3cm}|>{\centering\arraybackslash}p{3cm}|}
\hline
\multirow{2}{*}{References} & \multicolumn{2}{c|}{Temperature} \\\cline{2-3}
                         & BB MOT & SF MOT \\\hline
Cavity & $6.21\pm 0.44~\mu$K & $0.65\pm 0.02~\mu$K\\\hline
VSL & $9.33\pm 0.25~\mu$K & $0.71\pm 0.02~\mu$K\\\hline
Active stabilization & $6.82\pm 0.98~\mu$K & $0.68\pm 0.1~\mu$K\\\hline
\end{tabular}
\caption{${\rm ^{88}Sr}$ BB and SF MOT temperature using the cavity-stabilized laser system, comb-stabilized system with the 10~MHz VSL reference, and comb-stabilized system with active feedback from the MOT position.} 
\label{table:Temperature RF}
\end{table}

We also characterize a single-frequency (SF) MOT obtained using the three different laser locking methods. 
We create the SF MOT in a pulsed manner by ramping the range of the BB modulation of the MOT beams to zero, sweeping the center frequency to 700~kHz red-detuned from resonance, and simultaneously reducing the beam intensity to about 250~$\mu$W/cm$^2$.
We then further ramp the intensity down to 20~$\mu$W/cm$^2$.
During this process, the magnetic field gradient remains at 0.39\,G/cm.
This procedure allows atoms to be cooled to sub-\(\mu\)K temperatures.
Table~\ref{table:Temperature RF} lists the temperatures achieved in the BB and SF MOTs with the conventional laser system and the comb based system referenced using either method, measured using time of flight~\cite{TOF_Lett}.
We achieve temperatures below 10~$\mu$K in the BB MOT using all locking methods.
While the cavity-locked system and the comb-based system with MOT-position stabilization produce MOTs with very similar temperatures, we believe the comb-based system referenced to VSL produces a slightly higher temperature due to reduced polarization purity of the MOT beams when this data set was taken.
In all cases, we are able to achieve sub-$\mu$K temperatures in the SF MOT with around $6\times 10^7$ atoms, and we see comparable performance between the cavity-stabilized system and the comb-stabilized system. SF MOT temperatures depend on MOT density, MOT laser intensity and the short-term ($\sim1$\,ms) red MOT laser linewidth. The lowest temperature can only be achieved if the laser linewidth is narrower than the red MOT transition's natural linewidth of 7.5\,kHz. The lowest temperatures that have been reached are around 0.4\,$\mu$K for MOT laser intensities around $I_{\mathrm{sat}}$ \cite{Katori_Photon_Recoil_Temperature}. In all three cases we reach temperatures that are less than twice as large, on par with expectation for our MOT laser intensity of several $I_{\mathrm{sat}}$, indicating a sufficiently narrow linewidth of the MOT laser.

\begin{figure}[]
    \centering
    \includegraphics[width=8.5cm]{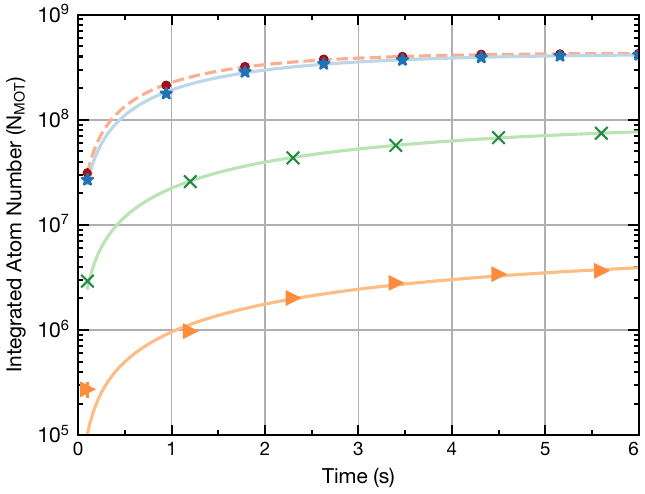}
    \caption[]
    {\label{fig: MOT load} Atom number over time during loading of a BB MOT for the three bosonic isotopes, $^{88}$Sr (blue stars), $^{86}$Sr (green crosses), and $^{84}$Sr (orange triangles), using the comb-stabilized laser system with the 10\,MHz reference from VSL. The loading data for the cavity-stabilized system and $^{88}$Sr is shown as a reference (red dots). All data is fitted to the following loading curve: $N(t)=N_{\mathrm{sat}}(1-e^{-t/\tau})$ and plotted with a solid (dashed) line for the comb (cavity)-stabilized system.
    }
\end{figure}

When using the comb-stabilized laser system, it is particularly easy to create MOTs of any of the bosonic Sr isotopes, as one can simply change the lock point of the red MOT laser. For all bosonic isotopes of strontium, we are able to load a continuous BB and a pulsed SF red MOT.
For the most abundant isotope, $^{88}$Sr, we achieve a maximum atom number of approximately $4.25\times10^8$ atoms with a loading time constant of 1.66~s with the VSL referenced frequency comb laser system, similar to the conventional laser system, shown in Fig.~\ref{fig: MOT load}.
As expected, the equilibrium atom number for the three isotopes is close to proportional to their natural abundance.

\subsection{QUASI-CONTINUOUS OUTCOUPLING FROM A MOT}
\label{sec:exp_setup:quasi-continuous}

\begin{figure*}[]
    \centering
    \includegraphics[scale=0.75]{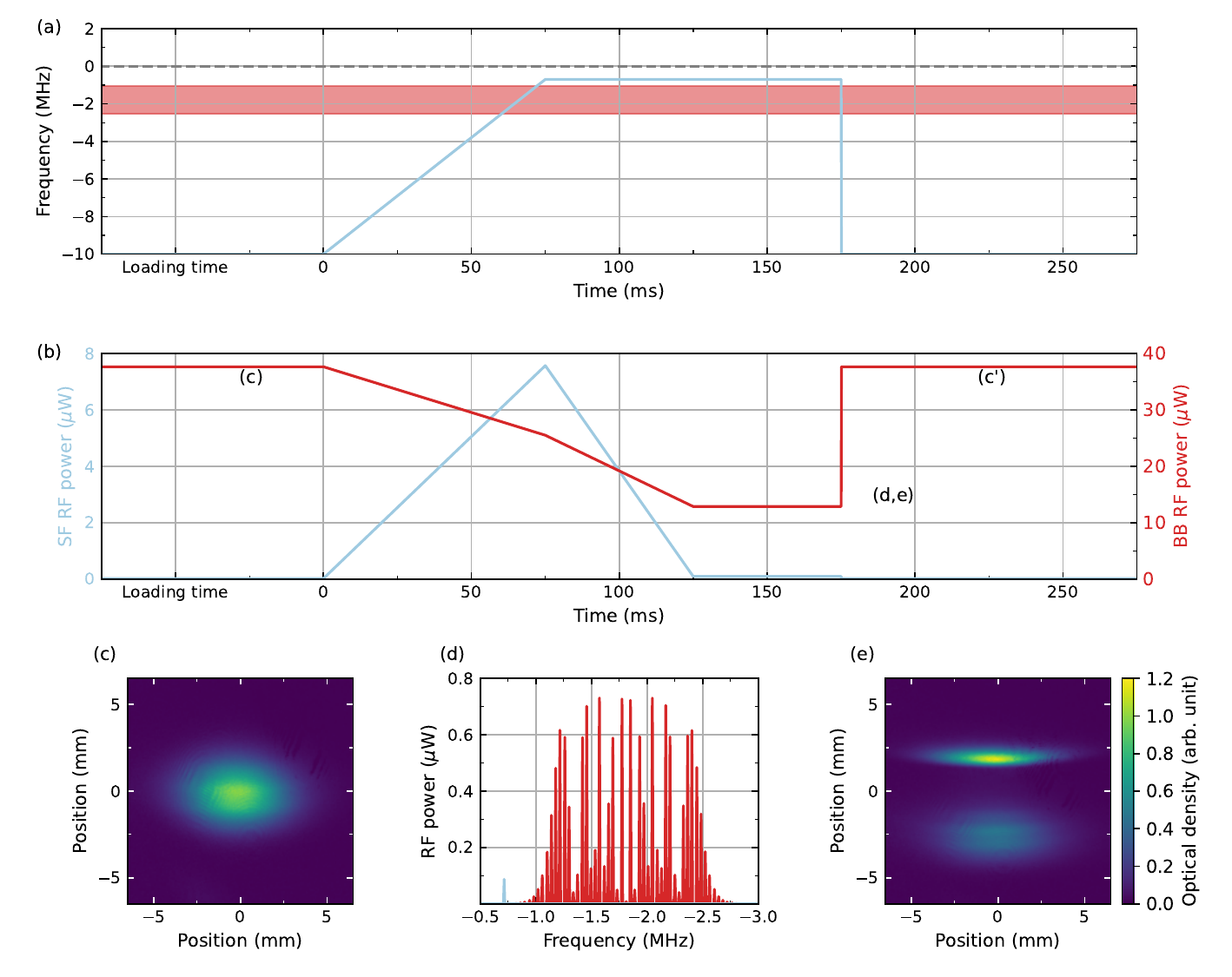}
    \caption{\label{fig:doublemot} Experimental sequence for quasi-continuous outcoupling of sub-$\mu$K cold atoms. (a,b) Frequency and power over time during outcoupling of atoms from the BB MOT (red) into the SF MOT (blue) to produce a dual-position MOT (e), for which N$_{\rm SF}=(42.9\pm1.3) \times 10^{6}$ atoms, N$_{\rm BB}=(66.2\pm4.5) \times 10^{6}$ atoms.
    The red MOT transition resonance frequency is denoted by the dashed black line in (a).
    (c) Absorption image of steady-state BB MOT.
    (d) RF Spectrum showing both BB (red) and SF (blue) tones for the dual-position MOT. SF tone is 700\,kHz red-detuned from resonance.
    Panels (c,d,e) correspond to the experimental stages with matching labels in panel (b).
    (c') Recapturing of atoms by switching off the SF MOT while keeping the BB MOT on. 
    Recapture probability from SF is $>90\%$:
    N$_{\rm with~SF}=(111.1\pm2.8) \times 10^{6}$ atoms and N$_{\rm without~SF}=(114.8\pm2.8) \times 10^{6}$ atoms.} 
\end{figure*}

A SF red MOT of strontium atoms is usually created in a pulsed fashion by ramping the MOT beam parameters over time, as described above. The laser trapping forces in such a MOT are weak compared to gravity, leading to a strong frequency dependence of the MOT position~\cite{MOT_pos_Ytterbium_Sillus,HeizenrederMOTSpec2025}. Exploiting this feature, we are able to simultaneously generate a BB and a SF MOT, located one below the other. An absorption image of this ``dual-position MOT” is shown in Panel~(e) of Fig.~\ref{fig:doublemot}. We create this dual-position MOT by simply adding an additional frequency tone to the MOT beams, as illustrated in Panel~(d) of Fig.~\ref{fig:doublemot}.

The red-shaded area in the top panel of Fig.~\ref{fig:doublemot} indicates the frequency range used for the BB MOT. The blue trace shows the evolution of the additional frequency tone used to create the SF component of the dual MOT. This frequency tone is ramped from 10\,MHz to 700\,kHz red-detuned from resonance over a duration of 75\,ms. During this ramp, the RF power, and thus the optical intensity of this tone, is increased to around $250~\mu\text{W}/\text{cm}^2$, while the intensity of the BB component is simultaneously decreased (see middle panel). This process enables a tunable fraction of atoms to be transferred into the SF MOT, controlled by the amplitude of the SF tone. 
The spatial position of the SF MOT, which ends more than 5\,mm away from the BB MOT center, is determined by the effective frequency difference between the starting frequency of the BB modulation and the SF tone. 
After the frequency ramp, the frequency is held constant for 50\,ms, while all frequency components are further reduced in intensity to optimize final cooling. We reach a SF intensity of around $20~\mu\text{W}/\text{cm}^2$. During this time, the blue MOT and Zeeman slower beams are switched off to avoid any unwanted heating from stray blue photons. These conditions are maintained for another 50\,ms before the SF component is switched off, the blue MOT and Zeeman slower beams are switched on again, and the BB component is restored to its original intensity to return to just the BB MOT.

This cycle of sweeping atoms from the BB MOT into the SF MOT can be repeated continually. The transfer rate is limited only by the loading rate of the BB MOT. At the end of each 175\,ms cycle, the gas cloud in the SF MOT reaches temperatures as low as 650\,nK. Approximately 90\% of the SF MOT atoms can be recaptured by the BB MOT at the beginning of the next cycle.

This high recapture efficiency makes our system ideally suited for fast cycling and quasi-continuous outcoupling of ultracold atoms into downstream traps, such as optical dipole traps, lattices, or tweezers, enabling experiments that require both high atomic densities and ultra-low temperatures.

When using the MOT position as a comb reference, we should be able to use the dual-position MOT technique to continuously monitor the fluorescence of the BB red MOT while ramping a SF tone with small intensity to create the SF red MOT.
We find that the MOT tracker is able to continue measuring the BB MOT position even while transferring some atoms to the SF MOT, and the low amount of fluorescence produced by the SF MOT only has a modest effect on the position measurement (see Appendix~\ref{App:a}).

\section{Conclusions and outlook}
\label{sec:conclusions}

We have demonstrated, to the best of our knowledge, the first quasi-continuous sub-\(\mu\)K strontium source achieved without the use of a free-space, high finesse cavity-stabilized laser. Our dispersion-engineered frequency comb enables sub-kHz linewidth stabilization by optimizing intracavity dispersion and implementing a robust PID-locking architecture. This eliminates the need for a high-finesse optical cavity and allows for direct laser locking onto the frequency comb.

Long-term frequency stability is ensured by synchronizing the comb’s repetition rate in the RF domain. We validate two stabilization approaches: one using a stable 10\,MHz reference from the Dutch Metrology Institute (VSL), and another using feedback from the MOT position with no external reference. These schemes are combined with a continuous ultra-cold strontium source housed in a compact vacuum system of approximately 0.5\,m\(^3\) volume, featuring spatially separated cooling and trapping regions. This architecture requires only two laser wavelengths, no repumping stages, and supports quasi-continuous operation of a MOT at sub-\(\mu\)K temperatures.

We achieve cooling performance comparable to traditional cavity-stabilized systems in terms of atom number, loading rate, and temperature across both BB and SF MOT stages addressing the narrow 7.5\,kHz intercombination line. Additionally, we introduce a technique for quasi-continuous atom transfer from the BB to the SF MOT using an additional frequency tone. For \({}^{88}\)Sr, this approach yields atomic samples as cold as 700\,nK. The resulting source is ideally suited for loading into optical lattices, dipole guides or tweezer arrays for downstream experiments such as continuously operating optical clocks.

The combination of a wide-band, fiber-based optical frequency comb with a compact, robust ultracold atom architecture offers a compelling path toward field-deployable quantum technologies. Owing to the comb’s broad spectral coverage, this approach can be extended beyond strontium to other alkaline-earth(-like) and lanthanide species, particularly those compatible with five-beam MOT configurations \cite{karim_single_atom_2025,five-beam_magneto-optical_trap_erbium_dysprosium}. The self-contained MOT-position stabilization scheme further reduces dependence on external RF references, enhancing robustness for operation in resource-limited environments. Together, these advances pave the way for high-performance, self-contained, continuously operating ultracold atom devices for applications in quantum sensing, precision timekeeping, and fundamental physics.

\begin{acknowledgments}

We thank Rob Smets for initiating the first White Rabbit implementations and laying the foundation for the White Rabbit service at SURF. We thank OPNT B.V. for setting up and calibrating the White Rabbit network that was used during our research. We thank N.J. van Druten, S. Wolzak and S. Sarkar for establishing the White Rabbit connection at UvA. We thank N.J. van Druten for thoughtful discussions and careful reading of the manuscript.

This work has received funding from the European Union’s (EU) Horizon 2020 research and innovation program under Grant Agreement No. 820404 (iqClock project) and No. 860579 (MoSaiQ project). It also received funding from the European Partnership on Metrology, co-financed from the European Union’s Horizon Europe Research and Innovation Programme and by the Participating States, under project 22IEM01 TOCK. It further received funding from the Dutch National Growth Fund (NGF), as part of the Quantum Delta NL programme.

The raw data and the analysis tools used in this manuscript can be found in Reference \cite{boughdachi2025data}.
\end{acknowledgments}

\appendix
\section{DUAL MOT LIFE TIME, TEMPERATURE AND LONG TERM STABILIZATION SCHEME}
\label{App:a}

\begin{figure}[]
    \centering
    \includegraphics[width=8.5cm]{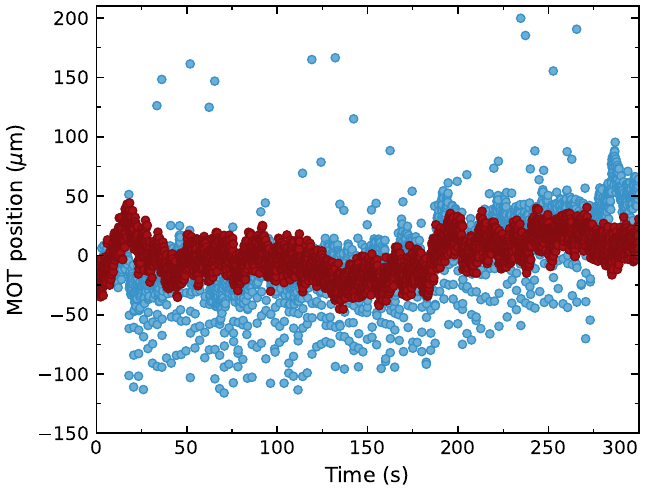}
    \caption[]
    {\label{fig: Trackerperformance} Position over 300~s of the BB MOT with no ramp to dual-position MOT (red dots) and with a ramp to dual-position MOT repeated 100 times (blue dots). The average atom number is comparable in both cases.
    The cavity-stabilized BB MOT position has a standard deviation of 15.5~$\mu$m, whereas with ramping to dual-position MOTs, the standard deviation increases to just 35.5~$\mu$m.
    Outlier points occur because the image taking is not synchronized with the ramps to the dual-position MOT. }
\end{figure}

In this section, we discuss details of the feedback loop that was used to stabilize the comb by tracking and stabilizing the MOT position.
The performance of the MOT tracker is shown in Fig.~\ref{fig: Trackerperformance} for two situations: using a BB MOT in steady-state, or continually ramping from BB to dual MOT.
First, we measured the position of the BB red MOT over 300~s, while the MOT tracker was actively measuring the position of the MOT.
This data set is shown by the dark red dots.
Next, we repeated this measurement, but simultaneously performed ramps to a dual-position MOT 100 times over the 300~s.  
This data set is shown by the blue dots. 
The MOT tracker fluorescence imaging has an image exposure time of 10\,ms and a rate of 40\,Hz, which is not synchronized with the dual-MOT ramps. This sometimes leads to outlier points when the image is taken while the SF MOT is close in position to the BB MOT, making it hard for the MOT tracker to precisely determine the BB MOT position.
However, this has only a modest effect on the performance of the MOT tracker performance, as the standard deviation in MOT position increases only from $15~\mu$m to $35.5~\mu$m as compared to the data run without ramping to SF. In the future, this effect could easily be avoided by synchronizing the MOT tracker image taking with the outcoupling ramps.

By taking an absorption image after a full cycle of loading the BB MOT, ramping to SF, and recapturing atoms in the BB MOT, we can measure the final atom number to determine the efficiency of the recapture process.
We find that around 90\% of SF MOT atoms are recaptured by the BB MOT after they are released from the SF MOT at the end of each ramp cycle.
This shows how the dual-position MOT can be used for near-continuous outcoupling of atoms from the red MOT, while still making use of the ease and stability of locking to the frequency comb.

\begin{figure}[]
    \centering
    \includegraphics[width=8.5cm]{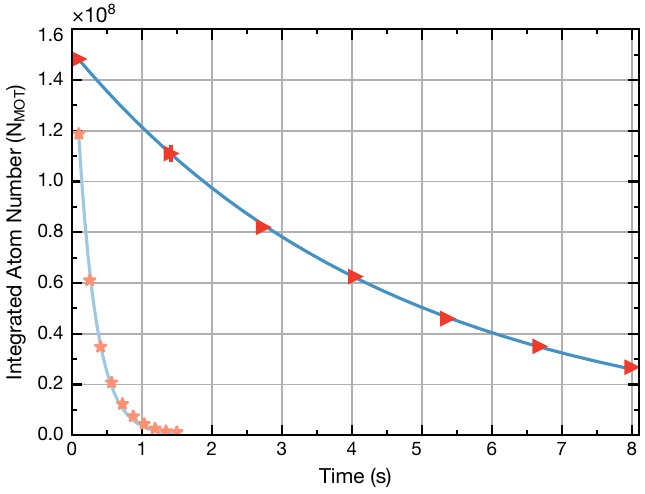}
    \caption[]
    {\label{fig:Life time} Integrated atom number vs. hold time for the BB MOT (red triangles) and the SF MOT (orange stars).
    To extract the lifetime we fit the data by$N(t) = N_{\mathrm{start}} e^{-t/\tau}$. The lifetime for the BB MOT is $\tau_{\mathrm{bb}} = 4.55\pm0.07$~s, and the lifetime of the SF MOT is $\tau_{sf} = 0.259\pm0.004$~s.}
\end{figure}

We further assess the performance of the comb-locked system by examining the lifetime of the BB and SF MOTs by measuring the atom number over a hold time.
The BB MOT has a lifetime of $\tau_{\mathrm{bb}} = 4.55\pm0.07$~s, which is comparable to the lifetime measured using a cavity-stabilized system.
The SF MOT has a shorter lifetime of $\tau_{\mathrm{sf}} = 0.259 \pm 0.004$~s. Due to the low laser intensities and the presence of only a single frequency component, the probability of photon absorption in the limited region where the Zeeman shift enables resonance is less than unity. This opens a significant downwards loss channel from the SF MOT~\cite{Katori_Photon_Recoil_Temperature}.
Additionally, we do not detect any difference in temperature between the SF MOT and the SF component of the dual-position MOT, which suggests that there is no significant effect of the BB MOT light on the SF lifetime in the dual MOT.
For a continuously or quasi-continuously operating machine this lifetime is sufficient, as atoms will be loaded into the next trap stage much faster.

\begin{table}[H]
\centering
\begin{tabular}{|>{\centering\arraybackslash}p{1.75cm}|>{\centering\arraybackslash}p{3cm}|>{\centering\arraybackslash}p{3cm}|}
\hline
\multirow{2}{*}{Isotope} & \multicolumn{2}{c|}{Temperature} \\\cline{2-3}
                         & BB MOT & SF MOT \\\hline
88 & $6.21\pm 0.44~\mu$K & $0.65\pm 0.02~\mu$K\\\hline
86 & $9.61\pm 0.48~\mu K$ & $0.70\pm 0.01~\mu K$\\\hline
84 & $9.03\pm 0.33~\mu K$ & $0.74\pm 0.08~\mu K$\\\hline
\end{tabular}
\caption{Temperatures measured for different isotopes using the cavity-locked system. The uncertainties are the standard deviation obtained from a fit to the atomic cloud width over expansion time after MOT switch-off.}
\label{table:Temperature isotopes}
\end{table}

In Table~\ref{table:Temperature isotopes}, we show the temperatures achieved for all three bosonic isotopes of strontium for the BB and SF MOTs with the cavity-locked system.
The temperatures achieved with the cavity-locked system are comparable to those achieved with the comb-stabilized system. We find that we achieve very similar temperatures for all isotopes, even though the scattering lengths are very different, making this machine versatile for a variety of experiments with strontium~\cite{Stellmer_84, Mickelson_scattering_88_86}.

\bibliography{main_V2}
\end{document}